# Brillouin Light Scattering Spectra as local Temperature Sensors for Thermal Magnons and Acoustic Phonons

Daniel R. Birt[1,2], Kyongmo An[2], Annie Weathers[3], Li Shi[1,3], Maxim Tsoi[1,2], and Xiaoqin Li[1,2,a]

1. Texas Materials Institute, The University of Texas at Austin, Austin, Texas 78712, USA
2. Department of Physics, Center of Complex Quantum Systems, The University of Texas at Austin, Austin, Texas 78712, USA
3. Department of Mechanical Engineering, The University of Texas at Austin, Austin, Texas 78712, USA

Email address: elaineli@physics.utexas.edu

## Abstract

We demonstrate the use of the micro-Brillouin light scattering (micro-BLS) technique as a local temperature sensor for magnons in a Permalloy (Py) thin film and phonons in the glass substrate. When the Py film is uniformly heated, we observe a systematic shift in the frequencies of two thermally excited perpendicular standing spin wave modes. Fitting the temperature dependent magnon spectra allows us to achieve a temperature resolution better than 2.5 K. In addition, we demonstrate that the micro-BLS spectra can be used to measure the local temperature of magnons and the relative temperature shift of phonons across a thermal gradient. Such local temperature sensors are useful for investigating spin caloritronic and thermal transport phenomena in general.

The discovery of the spin Seebeck effect has ushered in the emerging field of spin caloritronics.[1] The spin Seebeck effect results from the combination of two phenomena: a thermally driven spin current and the inverse spin Hall effect in a metal with strong spin-orbit coupling, such as Pt. The spin Seebeck effect or related spin-dependent Seebeck effect and spin-dependent Peltier effect have been observed in a variety of materials including magnetic metals,[2-4] insulators,[5] and ferromagnetic semiconductors.[6] In magnetic insulators such as yittrium iron garnet (YIG), the spin current is carried by spin waves or magnons. In magnetic metals such as Permalloy (Py, $Ni_{80}Fe_{20}$), the spin current can be carried by both conduction electrons and magnons. To understand the spin current driven by a thermal gradient, it is important to characterize the local temperatures of magnons and phonons.

Micro-Brillouin light scattering (micro-BLS) is a powerful tool for investigating thermal magnons.[7] It offers the typical advantages of optical spectroscopy, such as high spatial resolution and non-contact measurement. Even more important for the purpose of investigating thermal phenomena in magnetic materials, micro-BLS experiments can probe the thermally excited magnon spectra. This invaluable information is missing in the standard magneto-optic Kerr effect (MOKE) microscopy experiments. In this Letter, we demonstrate that the temperature-dependent thermal magnon spectra and acoustic phonon spectra may be used as local temperature sensors for investigating spin caloritronics and thermal transport phenomena.

Our experimental setup is illustrated in Fig. 1. The sample consists of a Py film sputtered on a glass substrate. The film is nominally 60 nm thick and coated with 10 nm of $SiO_x$ to prevent oxidation. The substrate is a rectangular piece of glass approximately 1 mm wide and 3 mm long. A magnetic field of 500 Oe was applied parallel to the short side of the sample. For the uniform heating measurements (Fig. 1(a)), the sample was mounted on a Peltier heater that was much larger than the sample. For the thermal gradient measurements (Fig. 1(b)), one end of the film was mounted on a resistive heater while the other end was mounted on a heat sink. A thermal gradient was applied along the length of the sample and perpendicular to the external magnetic field. A linearly polarized, single frequency laser at 532 nm was directed normal to the sample surface and focused to a spot size of ~1 μm. The average laser power applied was less than 2 mW to avoid overheating the Py film. The component of the backscattered light with orthogonal (parallel) polarization to the incoming beam was sent to a Sandercock-type multipass tandem Fabry-Pérot interferometer to resolve the inelastically scattered light from magnons (phonons). A computer controlled, imaging-based position correction algorithm ensured that the beam remained in focus throughout the measurement.

We first discuss how to interpret the micro-BLS magnon spectra, some examples of which are shown in Fig. 2(a). The spectra are dominated by two strong peaks corresponding to the first two perpendicular standing spin wave (PSSW) modes ($n = 1$, $n = 2$) of the Py film. The PSSW modes are characterized by sinusoidal amplitude profiles in the film thickness direction with the number of nodes corresponding to $n$, as illustrated in the inset of Fig. 2(a).[8] The magnetostatic surface wave ($n = 0$) mode is present but difficult to identify in the figure. It appears as a background surrounding the $n = 1$ peak due to its broad linewidth. The same range of wavevectors are probed for all spin wave modes,[9] which corresponds to different ranges of frequencies due to their different dispersion relations, leading to dissimilar linewidths in the BLS spectra. The linewidth of the $n = 0$ mode is approximately five times broader than that of the $n = 1$ mode according to the calculated dispersion relation.[8] For the rest of the paper, we focus on n = 1 and $n = 2$ modes because their narrow linewidths lend themselves as sensitive temperature sensors.

For our experiments with normal incident light, the frequencies of the PSSW modes can be calculated according to Eq. 1,[10]

$$\omega_n = \gamma\sqrt{\left(H_{ext} + \frac{2A}{M_S}\left(\frac{n\pi}{L}\right)^2\right)\left(H_{ext} + \frac{2A}{M_S}\left(\frac{n\pi}{L}\right)^2 + 4\pi M_S\right)} \qquad (1)$$

where $n$ is the number of nodes of the PSSW mode in the thickness direction, $\gamma$ is the gyromagnetic ratio (2.93 MHz/Oe), $A$ ($\sim 1.0\times10^{-6}$ erg/cm at $\sim$300 K)[11] is the exchange constant, $H_{ext}$ is the applied external field, $L$ is the thickness of the ferromagnetic film, and $M_S$ is the saturation magnetization. The temperature dependence of the exchange constant and saturation magnetization cause the mode frequency to vary with temperature. To confirm our interpretation of the magnon modes, we extract the peak positions from the spectra as the external field is varied. A few examples of the raw magnon spectra at different external fields are shown in Fig. 2(a). The peaks corresponding to the PSSW modes are fit with Lorentzian lineshapes, and the extracted peak positions are shown in Fig. 2(b). We fit the extracted values to Eq. 1 to obtain the saturation magnetization and the film thickness. The parameters obtained from the fitting are $L = 62.8 \pm 0.1$ nm and $4\pi M_S = 10{,}900 \pm 100$ Oe. The value for the saturation magnetization agrees well with those found in the literature for Py at room temperature.[12,13]

We then calibrated how the magnon frequencies shift as a function of temperature when the sample is heated uniformly, as shown in Fig. 3. The magnon frequency was extracted by fitting each PSSW peak in the BLS intensity with a Lorentzian lineshape, as for the magnetic field dependent measurements. At each temperature, 24 independent measurements (for a total measurement time of two hours) were taken. The measurements were fit individually and the centers of the peaks were averaged. The error in frequency quoted for each temperature is the standard deviation of the mean. Over the temperature range studied, the frequency decreases linearly with increasing temperature due to the reduced saturation magnetization and exchange constant at higher temperatures. Linear fits to the data yield slopes of $(-8.4 \pm 0.4) \times 10^{-3}$ GHz/K and $(-16.1 \pm 0.8) \times 10^{-3}$ GHz/K for the $n = 1$ and $n = 2$ modes, respectively. The average error in the frequencies for the two modes are 0.02 GHz for $n = 1$ and 0.05 GHz for $n = 2$. Therefore, the measured temperature sensitivities are about 2.4 K and 3.1 K for the $n = 1$ and $n = 2$ modes, respectively. The temperature sensitivities quoted in this study can be further improved simply by extending the measurement time, but the measured values already compare favorably with other nanoscale thermometry techniques.[14]

We now investigate the feasibility of using the BLS spectra as a local magnon temperature sensor. To do so, we establish an in-plane temperature gradient by mounting the sample to a heater and heat sink, as illustrated in Fig. 1(b). We measure the magnon spectra at different locations along the temperature gradient by moving the sample on a motorized stage in 10 steps of 160 µm each. We performed 24 independent measurements at each position, averaged the peak positions fitted with a Lorentzian lineshape, and quoted the standard deviation of the peak positions as the error bar. The frequency of the PSSW modes can be used as a temperature sensor for the local magnon temperature when compared to the frequency calibration performed on the uniformly heated film, as shown in Fig. 4. To account for the spatial variation of the magnon frequencies due to spatially non-uniform properties of the film, we subtracted the measured magnon frequencies at each location without the thermal gradient from those with the thermal gradient. The resulting magnon frequency shifts for the $n = 2$ and $n = 1$ PSSW modes are displayed in Figs. 4(a) and 4(b), respectively. The spatial dependence of the magnon spectra corresponds to a linear temperature profile, as expected in this simple geometry. We fit the frequency shifts to a linear function and obtain shifts of $(-8.9 \pm 1.3) \times 10^{-2}$ GHz/mm for the $n = 1$ mode and $(-20 \pm 2) \times 10^{-2}$ GHz/mm for the $n = 2$ mode. Using the uniform heating data, the measured frequency shifts correspond to a -10.6 ±1.2 K/mm temperature gradient between the hot and the cold ends for the $n = 1$ mode and -12.6 ±1.2 K/mm as measured with the $n = 2$ mode. We note that the temperature gradients extracted from the two PSSW modes are consistent.

To complement the magnon temperature measurement, we measure the acoustic phonon spectra of the glass substrate. The spectrum contains one peak corresponding to the frequency of the acoustic phonons in glass with wave vectors of $4\pi N/\lambda$, where $\lambda$ is the laser wavelength and $N$ is the index of refraction at the laser wavelength. Assuming $N = 1.5$, the measured frequency of 35 GHz implies a speed of sound of 6 km/s, which is quite reasonable for glass. We first calibrated the change in the phonon spectra as a function of temperature under uniform heating conditions with the setup shown in Fig. 1(a). Unlike the magnon spectra, the phonon frequency did not shift as the temperature was varied. Instead, a systematic increase in the intensity of the phonon peak was observed (Fig. 5(a)) as the temperature was raised. Because the ratio between the phonon energy and $kT$ is small, the Bose-Einstein distribution for the phonon mode is reduced to the classical limit that depends linearly on temperature, i.e., $\langle n \rangle = 1/(\exp(\hbar\omega/kT) - 1) \approx kT/\hbar\omega$. Consequently, the intensity of the phonon peak in the BLS spectra depends linearly on temperature, as shown in Fig. 5(b). The intensities plotted in Fig. 5(b) are the averaged values from 24 measurements with the standard deviation displayed as the error bar. A linear fit yields that the phonon intensity increased by 22 ±2 % as the temperature increased from 296 K to 360 K, also an increase of 22 %. The linear fit yields a slope of 0.35 ±0.03 %/K. To map out the local temperature of the sample with a thermal gradient, we subtracted the intensities at each point with and without the thermal gradient to compensate for spatial non-uniformities in optical properties of the substrate. This procedure allows us to obtain a relative temperature shift for the phonons along the temperature gradient. The data shown in Fig. 5(c) is fit with a linear function with a slope of 6.3 ±1.1 %/mm. Using the slope obtained from the uniform heating data, this change in BLS intensity of the phonon mode implies a temperature gradient of 18 ±3 K/mm over the region measured, which is somewhat larger than the value extracted from the magnon spectra but still within two standard deviations. The larger temperature change in the phonons may be partially explained by the fact that the heater is attached to the side of the substrate opposite from the Py film. Thus, a vertical thermal gradient may exist across the thickness of the substrate.

Our experiments demonstrate that the BLS spectra can be used as local temperature sensors for magnons and phonons. The frequency shift of magnon spectra provides an absolute temperature scale. In the case of the phonon, relative temperature changes were measured based on the changes in BLS intensity. We note that relative temperature change is of significant interest in measuring thermal properties. For example, a micro-Raman technique has been used to measure relative temperature change in suspended graphene sheets from which the thermal conductivity of graphene was determined.[15] Local magnon and phonon temperature sensors may be applied to investigate spin caloritronic and thermal transport phenomena. For example, a key assumption in the current understanding of thermally driven spin currents is that magnons temperature may not reach thermal equilibrium with the phonons locally. This difference between the phonon and magnon temperatures underlies the so called "phonon-magnon drag" that leads to the directional magnon transport or spin current in magnetic insulators and semiconductors.[16-19] Direct experimental confirmation of this phenomenon is currently lacking most likely due to the small temperature difference between these two temperatures in a sample with a modest temperature gradient.[20] Further improvement in the BLS measurement sensitivity of the magnon and phonon temperatures and the use of a large temperature gradient may help to clarify the fundamental mechanism responsible for thermally driven spin currents. In addition, to investigate thermal transport in a non-magnetic microstructure, a thin magnetic metal film may be deposited on the top of the microstructure and used as a local temperature sensor. Moreover, the phonon temperature obtained from the BLS spectra is that of the low-frequency acoustic phonons. In comparison, micro-Raman spectroscopy can be used to obtain the optical phonon temperature. Hence, the combination of these two techniques can be used to investigate highly non-equilibrium phonon transport, such as the hot optical phonon phenomena predicted in silicon and carbon nanotube-based electronic devices.[21,22]

**Acknowledgements**

We gratefully acknowledge financial support from the following sources: AFOSR FA9550-08-1-0463, AFOSR FA-9550-08-1-0058, and the Alfred P. Sloan Foundation. DB acknowledges a fellowship from the NSF-IGERT program via grant DGE-0549417. AW and LS acknowledge support from NSF Thermal Transport Processes Program (CBET-0933454) and a NSF Graduate Research Fellowship to AW.

**Figures**

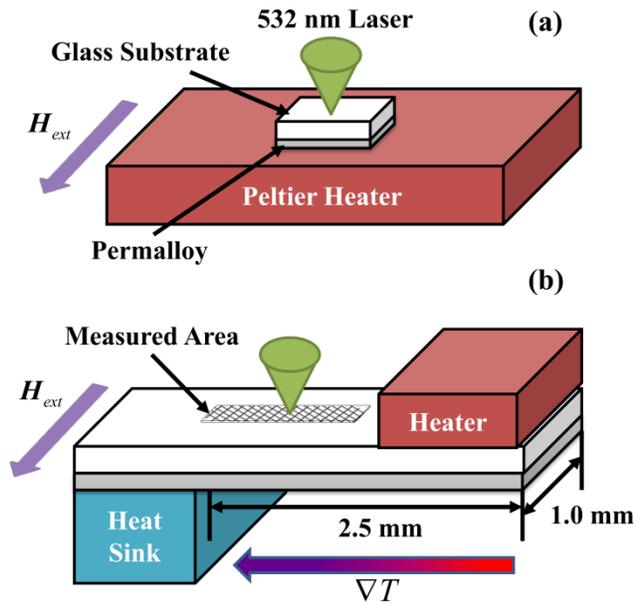

Fig. 1: Experimental schematics. (a) Uniform heating. (b) A thermal gradient is applied laterally to the substrate and perpendicular to the magnetic field.

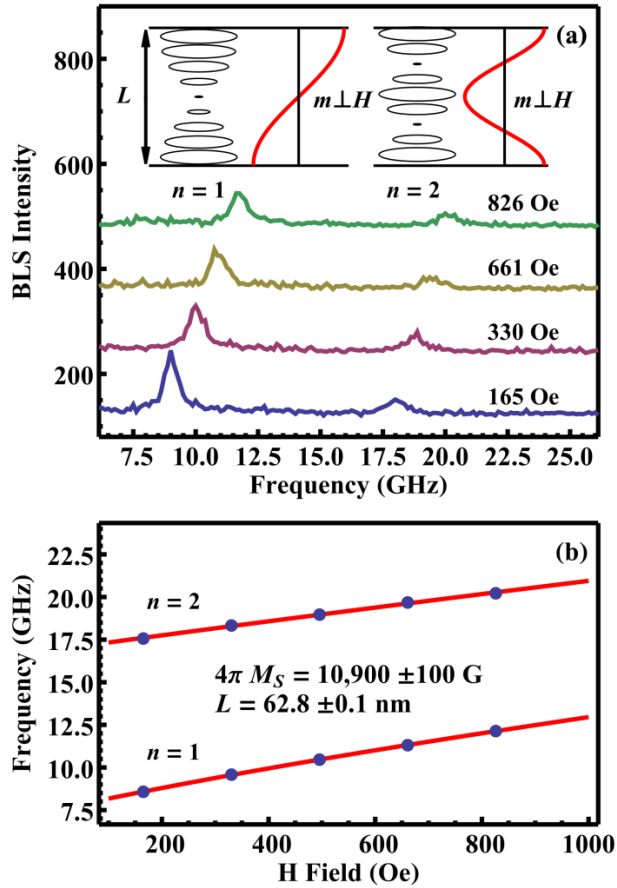

Fig. 2: Magnetic field dependent thermal magnon spectra. (a) Two PSSW modes are observed ($n = 1$ and $n = 2$ modes). The insets illustrate the behavior of the dynamic magnetization in the thickness direction of the film. The ellipses show the path traced out by the dynamic components of the magnetization, and the red sinusoidal curves represent the amplitudes of one of the dynamic components. (b) The frequencies of the modes were extracted as described in the text and fit with Eq. 1, which allows us to determine the saturation magnetization at room temperature.

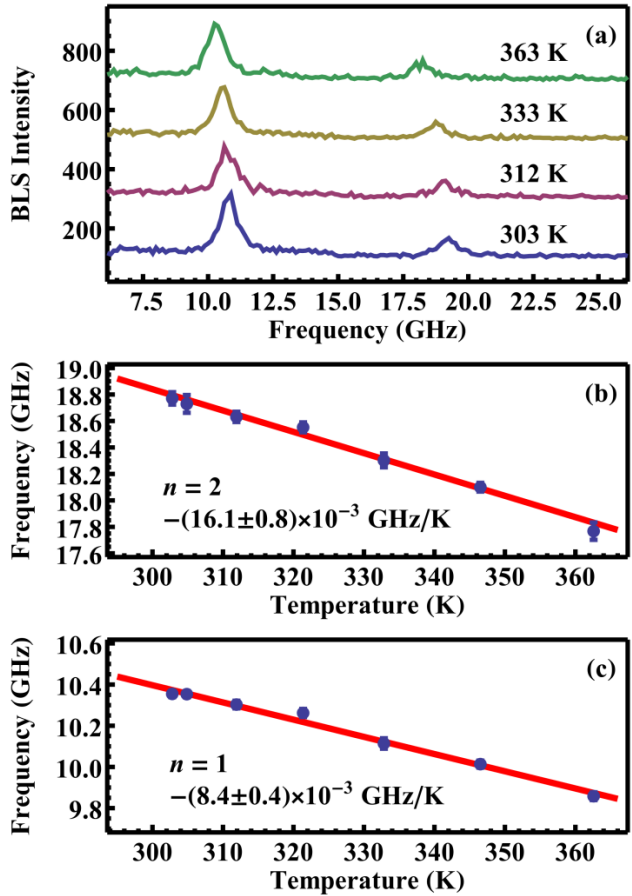

Fig. 3: Temperature dependent thermal magnon spectra in a uniformly heated film. (a) Thermal magnon spectra with increasing temperature from the bottom to the top. Extracted magnon frequencies as a function of temperature for the (b) $n = 1$ and (c) $n = 2$ modes. The solid lines are linear fits, which allow us to determine the frequency shift as a function of temperature.

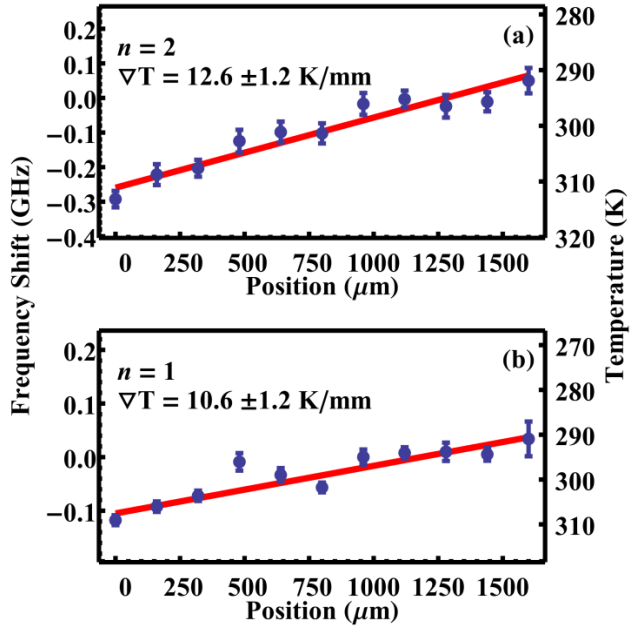

Fig. 4: The extracted magnon frequency shift and associated temperature for the (a) $n = 1$ and (b) $n = 2$ modes as the beam was scanned along the temperature gradient. The magnon frequency without a thermal gradient is subtracted from that with the thermal gradient to calculate the frequency shift, and a frequency shift of zero corresponds to room temperature. The temperature scale is obtained by comparing the magnon frequencies to those obtained in the uniformly heated film. The approximate linear shift in magnon frequency is consistent with a linearly varying temperature profile.

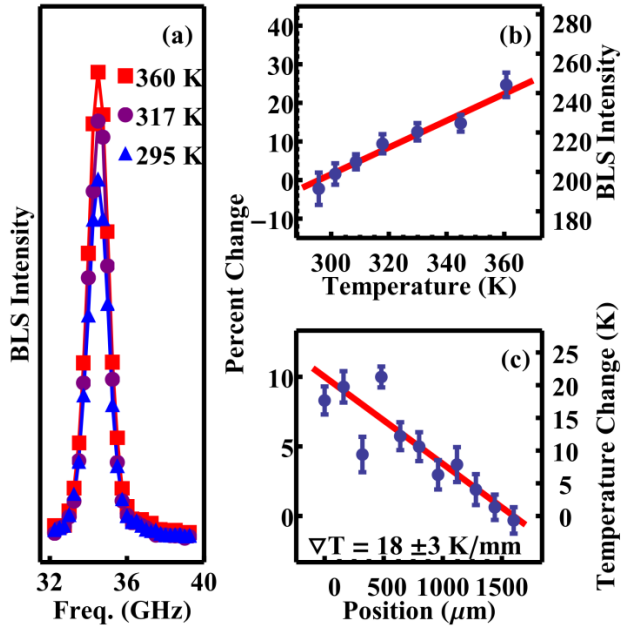

Fig. 5: (a) Phonon spectra of the glass substrate collected at three different temperatures under uniform heating, which shows how the peak intensity increases as the temperature increases. (b) Percent change of the intensity of the phonon peak as the temperature was increased under uniform heating. (c) Percent change of the intensity as the beam was scanned along the temperature gradient, which shows a linearly varying temperature profile. The change in temperature relative to that measured at the furthest point from the heater is shown on the right axis.